%% file: AJEMJanssen_215_arXiv-version_07042014.tex
\documentclass[12pt,a4paper]{article}
\usepackage{amsmath}
\usepackage{graphicx}
\newcommand{\osal}{\boldsymbol\alpha}
\newcommand{\osbe}{\boldsymbol\beta}
\newcommand{\osga}{\boldsymbol\gamma}
\newcommand{\osde}{\boldsymbol\delta}
\input{layout.tex}

\title{Zernike expansions of derivatives and Laplacians of the Zernike circle polynomials}
\author{A.J.E.M.\ Janssen \\
Department of Mathematics and Computer Science, \\
Eindhoven University of Technology, \\
P.O.\ Box 513, 5600 MB Eindhoven, The Netherlands. \\
E-mail a.j.e.m.janssen@tue.nl}
\date{}

\begin{document}
\maketitle
\mbox{} \\ \\ \\
\noindent
{\bf \large Abstract} \\[2mm]
The partial derivatives and Laplacians of the Zernike circle polynomials occur in various places in the literature on computational optics. In a number of cases, the expansion of these derivatives and Laplacians in the circle polynomials are required. For the first-order partial derivatives, analytic results are scattered in the literature, starting as early as 1942 in Nijboer's thesis and continuing until present day, with some emphasis on recursive computation schemes. A brief historic account of these results is given in the present paper. By choosing the unnormalized version of the circle polynomials, with exponential rather than trigonometric azimuthal dependence, and by a proper combination of the two partial derivatives, a concise form of the series expressions emerges. This form is appropriate for the formulation and solution of a model wave-front sensing problem of reconstructing a wave-front on the level of its expansion coefficients from (measurements of the expansion coefficients of) the partial derivatives. It turns out that the least-squares estimation problem arising here decouples per azimuthal order $m$, and per $m$ the generalized inverse solution assumes a concise analytic form, thereby avoiding SVD-decompositions. The preferred version of the circle polynomials, with proper combination of the partial derivatives, also leads to a concise analytic result for the Zernike expansion of the Laplacian of the circle polynomials. From these expansions, the properties of the Laplacian as a mapping from the space of circle polynomials of maximal degree $N$, as required in the study of the Neumann problem associated with the Transport-of-Intensity equation, can be read off within a single glance. Furthermore, the inverse of the Laplacian on this space is shown to have a concise analytic form. 
\noindent
\section{Introduction and overview} \label{sec1}
\begin{figure}[htb!]
\includegraphics[width=0.85\textwidth]{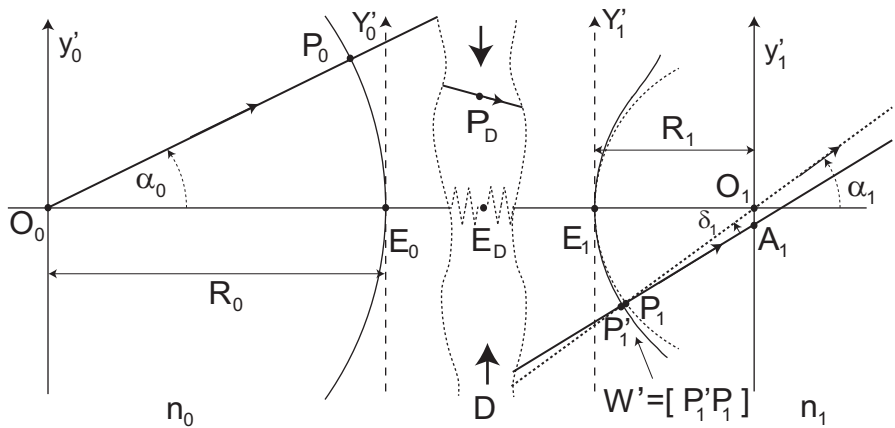}
\caption{A specific ray from the axial object point $O_0$ intersects the entrance pupil sphere through $E_0$ of an optical system in the point $P_0$ and the diaphragm in $P_D$. The aberrated ray intersects the exit pupil sphere through $E_1$ in $P_1$  and
the image plane in $A_1$. The wave-front belonging to the pencil from $O_0$ is the solid curve through $E_1$ and $P_1$. }
\label{Fig_01}
\end{figure}

%
\hspace*{6mm}The design and analysis of complex optical imaging systems is commonly carried out with the aid of ray tracing. To obtain information about the imaging quality of an optical system, a certain number of pencil of rays in the object plane is defined and the rays of each pencil are traced from each object point to the diaphragm of the optical system, see Fig.~\ref{Fig_01}. The rays that intersect the open area of the diaphragm are followed further through the interior of the optical system and the point of intersection with the exit pupil sphere is determined. By keeping track of the optical pathlength along each ray, the difference in pathlength $W^{'}=[O_0P_1]-[O_0E_1]$, denoted by $[P_1^{'}P_1]$ in the figure, can be established for each ray with respect to a certain reference ray. In  general, this is the ray $O_0E_0E_DE_1$ that passes through the axial point $ E_D$ of the diaphragm $D$.
As a result of the tracing of a large number of rays belonging to a particular object point, the wave-front in the exit pupil can be computed, for instance, by interpolation. For optical systems with wave-front deviations from a spherical surface  that are large with respect to the wavelength, the optical disturbance in the image plane is calculated by tracing rays beyond the exit pupil and calculating the intersection point $A_1$ with the image plane. In this way, the spot diagram is obtained. For the calculation of the ray directions from the wave-front surface data only, the gradient vector of the wave-front is needed. Following \cite{ref1} and referring to Fig.~\ref{Fig_01}, we perform the coordinate transformations
\begin{eqnarray}
&&X_1=X_1^{'}/\rho_0~, \;\;\;\;\;\;\;\;\;\;\;\;\;\;Y_1=Y_1^{'}/\rho_0~, \nonumber \\
&&x_1=x_1^{'}~\frac{n_1\sin\alpha_1}{\lambda_0}~, \;\;\;\;\;\;
y_1=y_1^{'}~\frac{n_1\sin\alpha_1}{\lambda_0}~, \nonumber \\
&&W=W^{'}/\lambda_0~,
 \label{Eq:01}
\end{eqnarray}
with $\lambda_0$ the vacuum wavelength and $n_1$ the refractive index in image space. The lateral  pupil coordinates on the exit pupil sphere have been normalized with respect to the lateral diameter of the pencil of rays on the exit pupil sphere. The image plane coordinates have been normalized with respect to the diffraction unit in the image plane and the wave-front aberration is expressed in units of the vacuum wavelength. Using these transformed pupil and field coordinates, the transverse aberration $O_1A_1$ in the image plane follows from
\begin{equation}
\delta x_1=\frac{\partial W(X_1,Y_1)}{\partial X_1}~, \;\;\;\;\;\;
\delta y_1=\frac{\partial W(X_1,Y_1)}{\partial Y_1}~,
\label{Eq:02}
\end{equation}
where we have used the approximation $P_1A_1=R_1$ that is valid for modest values of the angular aberration $\delta$. In most practical optical imaging systems, $\delta$ will not exceed 1 mrad.
For a general off-axis point, the entrance and exit pupil spheres are tilted such that their midpoints coincide with the paraxial object and image points. \\
\hspace*{6mm}Eq.(\ref{Eq:02}) shows that the gradient of the wave-front surface immediately yields the transverse aberration components in cartesian coordinates. Using normalized polar coordinates $(\rho,\theta)$ on the exit pupil sphere, we have
\begin{eqnarray}
\delta x_1&=& \cos \theta ~\frac{\partial W(\rho,\theta)}{\partial \rho}~ - ~
\frac{\sin \theta}{\rho}~ \frac{\partial W(\rho,\theta)}{\partial \theta}~,
\nonumber \\
\delta y_1&=& \sin \theta ~\frac{\partial W(\rho,\theta)}{\partial \rho} ~ + ~
\frac{\cos \theta}{\rho}~ \frac{\partial W(\rho,\theta)}{\partial \theta}~,
\label{Eq:03}
\end{eqnarray}
or, in complex notation,
\begin{equation}
\left (\begin{array}{c} \delta x_1 \\ \delta y_1 \end{array} \right )=
\Re \left \{ \exp (i\theta ) \left [ \frac{\partial W}{\partial \rho} + \frac{i}{\rho}~\frac{\partial W}{\partial \theta} \right ]
\left (\begin{array}{c} 1 \\ -i \end{array} \right ) \right \}~.
\label{Eq:03a}
\end{equation}
If it is desirable, Eq.(\ref{Eq:03a}) can be used equally well for complex variables; this is useful when $W$, the wave-front deviation in the exit pupil, is replaced by the complex amplitude $A(X,Y)\exp\{ikW(X,Y)\}$ on the exit pupil sphere.\\
\hspace*{6mm}In a practical situation, with a more or less circular cross-section of a pencil of rays or a propagating wave, the expansion of the wave-front aberration with polynomials that are orthogonal on the unit circle is appropriate (Zernike polynomials). Polynomials defined on the exit pupil sphere that allow an orthogonal Zernike expansion of the transverse aberration components $(\delta x_1, \delta y_1)$ have been proposed by Lukosz \cite{ref2}, but these polynomials are not strictly orthogonal on the exit pupil. \\
\hspace*{6mm}Other computational problems can be imagined in which wave gradients  are needed.  For instance, in electromagnetic problems, to calculate the energy and momentum flow, the wave normal has to be calculated over the area of the wave-front \cite{ref3}. In the case of adaptive optics wave-front correction, the local slope components of a wave-front are measured by means of a Shack-Hartmann sensor \cite{ref4}. In such a measurement problem, the gradient components of a wave-front, sampled in a sufficiently large number of points, have to be integrated to obtain the ideal wave-front. The correction of turbulence in the atmosphere during stellar observation is an example of such a combined measurement and computational task. Fast and efficient algorithms are needed because of the high temporal bandwidth of the turbulence effects \cite{ref5}.\\
\hspace*{6mm}Higher-order derivatives of the wave-front function are used to improve the reliability of the measurement data. For instance, the Euler principal curvatures and the azimuths  of the two corresponding principal planes are used in an enhanced wave-front reconstruction method for adaptive optics \cite{ref6}.
Higher-order derivatives of the wave-front function $W$ are also needed when approximate solutions of Maxwell's equations are pursued in free space \cite{ref7}. For field solutions that are valid far away from diffracting obstacles, at least in terms of the wavelength of the light,  an approximate solutions $U(x,y,z;t)$ = $Q(x,y,z)\exp\{i(kz-\omega t)\}$ is  used, with $Q(x,y,z)$ = $A(x,y,z) \exp \{ikW(x,y,z)\}$. It should be a slowly varying solution of the (paraxial) wave equation,
\begin{equation}
\nabla_t^2 Q + 2ik \frac{\partial Q}{\partial z} + k_2 Q = 0~,
\label{Eq:04}
\end{equation}
where the subscript $t$ means that only the derivatives with respect to the transverse coordinates $(x,y)$ need to be considered. The wave number $k$ equals $n~k_0$ with $k_0$ the wave number in vacuum of the radiation.\\
\hspace*{6mm}The intensity function $I=AA^{\ast}$ can be shown to satisfy the intensity transport equation \cite{ref7}-\cite{ref8},
\begin{equation}
-k \frac{\partial I}{\partial z} = \nabla_t \cdot [I\nabla_t (kW)] =
I\nabla_t^2 (kW) + \nabla_t I \cdot \nabla_t(kW)~,
\label{Eq:05}
\end{equation}
in which second order derivatives of the wave-front function $W$ need to be made available.\\
\mbox{ }
In this paper, we present analytic results for the gradients and Laplacians of the Zernike circle polynomials $Z_n^m$, in the unnormalized version with exponential azimuthal dependence, into which any wave-front function $W$ can be thought to be expanded. These results are useful in formulating and solving the reconstruction problem of wave-front functions from their first-order partial derivatives, and for tracking the solution of the Transport-of-Intensity equation, on the level of Zernike expansion coefficients. Choosing normalized Cartesian coordinates $\nu$, $\mu$ on the unit disk $\nu^2+\mu^2\leq1$, it turns out to be convenient to combine the partial derivatives $\frac{\partial W}{\partial\nu}$, $\frac{\partial W}{\partial\mu}$ according to $\frac{\partial W}{\partial\nu}+i\,\frac{\partial W}{\partial\mu}$ and $\frac{\partial W}{\partial\nu}-i\,\frac{\partial W}{\partial\mu}$, in accordance with two formulas for derivatives of the circle polynomials in polar coordinates as presented by Lukosz in \cite{ref2}, Anhang~II. Assuming to have available (measurements of the) Zernike coefficients of the two combinations $\frac{\partial W}{\partial\nu}\pm i\,\frac{\partial W}{\partial\mu}$, the least-squares problem of estimating the Zernike coefficients of the wave-front function itself has a very tractable form: The problem decouples per $m$, and, for any $m$, the pseudo-inverse solution $(A^HA)^{-1}A^H{\bf c}$ can be computed explicitly. Secondly, the Laplacian $\Delta=\nabla^2$ can be written as $(\frac{\partial}{\partial\nu}+i\,\frac{\partial}{\partial\mu})(\frac{\partial}{\partial\nu}-i\, \frac{\partial}{\partial\mu})$, and this yields a very concise and explicit formula for the Laplacian of the Zernike circle polynomials, and, hence, for any wave-front function $W$ developed into the circle polynomials. Moreover, $\Delta^{-1}Z_n^m$ can be shown to be an explicit linear combination of 3 circle polynomials of azimuthal order $m$.

The paper is organized as follows. In Section~\ref{sec2}, we present the basic formulas concerning the first-order partial derivatives of Zernike circle polynomials and first-order derivatives of the radial polynomials. As said, we use here the Born-and-Wolf convention \cite{ref3} with upper index $m$ the azimuthal order and lower index $n$ the degree, and with exponential azimuthal dependence $\exp(im\vart)$. We also give in Section~\ref{sec2} an account of the development of the results on first-order partial derivatives, and we present relations with Lukosz polynomials, pupil scaling and decentralization and generalized Zernike functions. We also make some comments on computation of the partial derivatives in Section~\ref{sec2}. In Section~\ref{sec3}, we consider the problem of estimating the Zernike coefficients of a wave-front function $W$ from (estimates of) the Zernike coefficients of the partial derivatives of first order of $W$. In Section~\ref{sec4}, we compute the Laplacians and the inverse Laplacians of the circle polynomials, and we make a connection with the work of Gureyev et al.\ in \cite{ref8} on the Neumann problem associated with the Transport-of-Intensity equation. In Appendix~A, we present the proofs of the results that give the action of the operators $\frac{\partial}{\partial\nu}\pm i\,\frac{\partial}{\partial\mu}$ in terms of the Zernike coefficients of a wave-front function $W$, the associated adjoint operators, and the generalized inverse required in the estimation problem of Section~\ref{sec3}. In Appendix~B, we present results on the inversion of certain special matrices that give rise to the explicit results in Sections~\ref{sec3}, \ref{sec4} on the inverse operators.

\section{First-order derivatives of the circle polynomials} \label{sec2}

For integer $n$ and $m$ with $n-|m|$ even and non-negative, we let, permitting ourselves some notational liberty,
\beq \label{e1}
Z_n^m(\nu,\mu)=Z_n^m(\rho,\vart)=R_n^{|m|}(\rho)\,e^{im\vart}~,
\eq
where we have set for real $\nu$, $\mu$ with $\nu^2+\mu^2\leq1$
\beq \label{e2}
\nu+i\mu=\rho\exp(i\vart)~;~~~~~~\nu=\rho\cos\vart\,,~~\mu=\rho\sin\vart~,
\eq
and where the radial polynomials $R_n^{|m|}$ are given as
\beq \label{e3}
R_n^{|m|}(\rho)=P_{\frac{n-|m|}{2}}^{(0,|m|)}(2\rho^2-1)
\eq
with $P_k^{(\alpha,\beta)}(x)$ the Jacobi polynomial of degree $k$ corresponding to the weight function $(1-x)^{\alpha}(1+x)^{\beta}$ on $[{-}1,1]$. We set, furthermore, $Z_n^m\equiv0$ for all integer $n$ and $m$, $n-|m|$ odd or negative.

\subsection{Basic identities for first-order derivatives} \label{subsec2.1}
\mbox{} \\[-9mm]

Let $n$ and $m$ be integers with $n-|m|$ even and non-negative. It follows from basic considerations about product functions as in Eq.~(\ref{e1}) in Cartesian and polar coordinates that
\begin{eqnarray} \label{e4}
& \mbox{} & \Bigl(\frac{\partial}{\partial\nu}\pm i\,\frac{\partial}{\partial\mu}\Bigr)\,Z_n^m(\nu,\mu) \nonumber \\[3mm]
& & =~\Bigl(\frac{\partial}{\partial\nu}\pm i\,\frac{\partial}{\partial\mu}\Bigr)\,[R_n^{|m|}((\nu^2+ \mu^2)^{1/2})\,\exp[im\arctan(\mu/\nu)]] \nonumber \\[3mm]
& & =~((R_n^{|m|})'(\rho)\mp\frac{m}{\rho}\,R_n^{|m|}(\rho))\,\exp[i(m\pm 1)\vart]~,
\end{eqnarray}
where the $'$ on the last line of Eq.~(\ref{e4}) denotes differentiation with respect to $\rho$. Next, see Subsection~\ref{subsec2.2} for comments, there is the identity
\beq \label{e5}
\Bigl(\frac{d}{d\rho}\pm\frac{m}{\rho}\Bigr)(R_n^{|m|}-R_{n-2}^{|m|})=2n\,R_{n-1}^{|m\mp 1|}~,
\eq
where we recall the conventions of the beginning of this section. Using Eq.~(\ref{e5}) with $n-2l$ where $l=0,1,...,\frac12(n-1-|m\mp1|)$, instead of $n$, summing over $l$, and using the observation that
\beq \label{e6}
\Bigl(\frac{d}{d\rho}\pm\frac{m}{\rho}\Bigr)\,R_{n-2-(n-1-|m\mp 1|)}^{|m|}= \Bigl(\frac{d}{d\rho}\pm\frac{m}{\rho}\Bigr)\,R_{|m\mp 1|-1}^{|m|}=0
\eq
in all cases, we get
\beq \label{e7}
\Bigl(\frac{d}{d\rho}\pm\frac{m}{\rho}\Bigr)\,R_n^{|m|}=2\,\sum_{l=0}^{\tfrac12 (n-1-|m\mp1|)}\,R_{n-1-2l}^{|m\mp1|}~.
\eq
It is sometimes convenient to realize that, in all cases, we can replace the upper summation limit in the series at the right-hand side of Eq.~(\ref{e7}) by $\frac12 (n-|m|)$ since all terms additionally included or excluded vanish by our conventions. We will also write the right-hand side of Eq.~(\ref{e7}) as
\beq \label{e8}
2\,\sum_{n'=|m\mp 1|(2)(n-1)}\,R_{n'}^{|m\mp1|}~,
\eq
where for integer $j$, $i$ with $j-i$ even and non-negative
\beq \label{e9}
k=i(2)j~~{\rm means}~~k=i,i+2,...,j~.
\eq
We get from Eqs.~(\ref{e4}) and (\ref{e7}), using in the latter upper summation limit $\frac12(n-|m|)$ as explained,
\beq \label{e10}
\Bigl(\frac{\partial}{\partial\nu}\pm i\,\frac{\partial}{\partial\mu}\Bigr)\,Z_n^m(\nu,\mu)=2\,
\sum_{l=0}^{\frac12(n-|m|)}\,(n-2l)\,Z_{n-1-2l}^{m\pm1}~.
\eq
Adding and subtracting the two identities in Eq.~(\ref{e10}), we get
\beq \label{e11}
\frac{\partial Z_n^m}{\partial\nu}=\sum_{l=0}^{\frac12(n-|m|)}\,(n-2l)\,Z_{n-1-2l}^{m-1}+\sum_{l=0} ^{\frac12(n-|m|)}\,(n-2l)\, Z_{n-1-2l}^{m+1}~,
\eq
and
\beq \label{e12}
\frac{\partial Z_n^m}{\partial\mu}=i\,\sum_{l=0}^{\frac12(n-|m|)}\, (n-2l)\,Z_{n-1-2l}^{m-1}-i\,\sum_{l=0}^{\frac12(n-|m|)}\, (n-2l)\, Z_{n-1-2l}^{m+1}~.
\eq
Alternatively, from Eqs.~(\ref{e4}) and (\ref{e5}), we get the recursive relation
\beq \label{e13}
\Bigl(\frac{\partial}{\partial\nu}\pm i\,\frac{\partial}{\partial\mu}\Bigr)\,Z_n^m= \Bigl(\frac{\partial}{\partial\nu}\pm i\,\frac{\partial}{\partial\mu}\Bigr)\,Z_{n-2}^m+2n\,Z_{n-1}^{m\pm1}~,
\eq
from which the recursions
\beq \label{e14}
\frac{\partial Z_n^m}{\partial\nu}=\frac{\partial Z_{n-2}^m}{\partial\nu}+n(Z_{n-1}^{m-1}+Z_{n-1}^{m+1})
\eq
and
\beq \label{e15}
\frac{\partial Z_n^m}{\partial\mu}=\frac{\partial Z_{n-2}^m}{\partial\mu}+in(Z_{n-1}^{m-1}-Z_{n-1}^{m+1})
\eq
for the separate first-order partial derivatives follow.

\subsection{History of the basic identities} \label{subsec2.2}
\mbox{} \\[-9mm]

The basic identities in Subsection~\ref{subsec2.1} have been (re)discovered, in one or another form, at various places in the optics literature from 1942 onwards. We give here a brief historic survey of what we have found in this respect. While the basic identities, as presented in Subsection~\ref{subsec2.1}, are given for general integer $m$, almost all writers on the subject make the restriction $m=0,1,...\,$. Showing validity of these identities for general $m$ from validity for $m=0,1,...$ is straightforward but requires some care.

The series identity in Eq.~(\ref{e7}) with $m\geq1$ and the $+$-sign has been given in Nijboer's 1942 thesis \cite{ref11} as Eq.~(2, 31), where the proof uses methods from complex function theory. The two identities in Eq.~(\ref{e5}) have been given in 1962 by Lukosz in \cite{ref2}, Anhang~II, Eqs.~(AII.4a,b) for the case that $m\geq0$, using the monomial representation
\beq \label{e16}
R_n^{|m|}(\rho)=\sum_{s=0}^{\frac12(n-|m|)}
\Bigl(\!\ba{c} n-s \\ \tfrac12(n-|m|) \ea\!\Bigr)\,\Bigl(\!\ba{c} \tfrac12(n-|m|) \\ s \ea\!\Bigr)\,\rho^{n-2s}
\eq
of the radial polynomials. Lukosz writes $m\mp1$ instead of $|m\mp1|$ at the right-hand side of Eq.~(\ref{e5}), which makes the result as presented by him somewhat doubtful in case that $m=0$. The recursion
\beq \label{e17}
\frac{d}{d\rho}\,(R_n^{|m|}-R_{n-2}^{|m|})=n(R_{n-1}^{|m-1|}+R_{n-2}^{|m+1|})~,
\eq
that follows from Eq.~(\ref{e5}) by adding the $\pm$-cases, has been presented for the case $m\geq0$ in 1976 by Noll in \cite{ref5} as Eq.~(13). Noll uses the integral representation
\beq \label{e18}
R_n^{|m|}(\rho)=({-}1)^{\frac{n-m}{2}}\,\il_0^{\infty}\,J_{n+1}(t)\,J_m(\rho t)\,dt~,~~~~~~ 0\leq\rho<1~,
\eq
also see \cite{ref12}, item 10.22.56, of the radial polynomials, together with recursion formulas for Bessel functions and their derivatives. The two series equations in Eq.~(\ref{e7}) have been presented for the case that $m\geq0$ in 1987 by Braat as \cite{ref13}, Eqs.~(10a,b), using Nijboer's result \cite{ref11}, Eq.~(2, 31) for the $+$-case and an argument based on the monomial representation in Eq.~(\ref{e16}) for the $-$-case (with a somewhat doubtful result for the case $m=0$). In Braat's forthcoming book \cite{ref14}, the two series identities in Eq.~(\ref{e7}) are proved by establishing Eq.~(\ref{e5}) via the integral result in Eq.~(\ref{e18}), using recursions for Bessel functions, and an induction step to go from Eq.~(\ref{e5}) to Eq.~(\ref{e7}). The identities in Eqs.~(\ref{e14}--\ref{e15}) have been given in 1999 by Capozzoli as \cite{ref15}, Eqs.~(10, 12) for $m\geq0$ and the versions of the circle polynomials with both the exponential and trigonometric dependence on $\vart$ (the case $m=0$ again being somewhat doubtful). The series representations in Eqs.~(\ref{e11}--\ref{e12}), for normalized circle polynomials with trigonometric azimuthal dependence, have been given in 2009 by the American National Standards Institute (ANSI) as \cite{ref15a}, Eqs.~(B.8--9). A similar thing has been done in 2014 by Stephenson in \cite{ref16}, Eqs.~(30--31). Both ANSI and Stepenson base their derivation on the recursion in Eq.~(\ref{e17}) that they ascribe to Noll.

\subsection{Relation with Lukosz polynomials, pupil scaling and generalized Zernike functions} \label{subsec2.3}
\mbox{} \\[-9mm]

In this subsection, we present some identities concerning the radial polynomials that follow from the basic identities in Subsec.~\ref{subsec2.1} and that connect these with various other issues in the theory of wave-front description using the circle polynomials.

The polynomials, see Eq.~(\ref{e5}),
\beq \label{e19}
L_n^m(\rho)=R_n^m(\rho)-R_{n-2}^m(\rho)~,~~~~~~ m=0,1,...\,,~~n=m+2,m+4,...
\eq
are called Lukosz polynomials and are considered in \cite{ref13}, see Eq.~(19) in \cite{ref13} where $B_n^m$ rather than $L_n^m$ is written, and \cite{ref14}, in connection with the study of transversal aberrations. The Lukosz polynomials are non-orthogonal with respect to the standard weight function $\rho\,d\rho$ on $[0,1]$. However, they are related to the generalized Zernike functions $R_n^{m,\alpha}$ as considered in \cite{ref17} according to
\beq \label{e20}
R_n^{m,1}={-}\,\frac{n-m+2}{2(n+2)}\,L_{n+2}^m~,~~~~~~m=0,1,...\,,~~n=m,m+2,...
\eq
that are orthogonal with respect to the weight function $(1-\rho^2)^{-1}\,\rho\,d\rho$ on $[0,1]$ according to
\beq \label{e21}
\il_0^1\,(1-\rho^2)^{-1}\,R_{n_1}^{m,1}(\rho)\,R_{n_2}^{m,1}(\rho)\,\rho\,d\rho= \frac{n-m+2}{(n+m+2)(n+2)}\,\delta_{n_1n_2}~,
\eq
with $n=n_1=n_2$ at the right-hand side of Eq.~(\ref{e21}).

Next, by subtracting the two cases in (\ref{e5}), we get the identity
\beq \label{e22}
\frac{m}{\rho}\,(R_n^{|m|}-R_{n-2}^{|m|})=n(R_{n-1}^{|m-1|}-R_{n-1}^{|m+1|})~.
\eq
In \cite{ref18}, \cite{ref19} optical systems with a variable numerical aperture are considered, and a central role is played by the identity
\begin{eqnarray} \label{e23}
R_{n'}^m(\eps\rho) & = & \sum_{n=m(2)n'}\,(R_{n'}^n(\eps)-R_{n'}^{n+2}(\eps))\,R_n^m(\rho) \nonumber \\[3mm]
& = & \sum_{n=m(2)n'}\,\frac{n+1}{n'+1}~\frac{1}{\eps}\,(R_{n'+1}^{n+1}(\eps)-R_{n'-1}^{n+1}(\eps)) \,R_n^m(\rho)
\end{eqnarray}
that yields an explicit Zernike expansion of a scaled radial polynomial $R_{n'}^m(\eps\rho)$ in terms of the unscaled radial polynomials $R_n^m(\rho)$. The identity between the two sets of coefficients required in Eq.~(\ref{e23}) is a consequence of Eq.~(\ref{e22}). In Eq.~(\ref{e23}) we have non-negative integers $n'$, $m$ with $n'-m$ even and non-negative, and the short hand notation $i(2)j$ of Eq.~(\ref{e9}) has been used.

\subsection{Computational issues} \label{subsec2.4}
\mbox{} \\[-9mm]

We make a few comments on the computation of radial polynomials and their derivatives. The direct representation of the radial polynomials, as given in Eq.~(\ref{e16}), can only be used for low values of $n$. There are, however, nowadays several good methods for computing (linear combinations of) radial polynomials of arbitrarily large degree $n$, see \cite{ref20}, \cite{ref21}, \cite{ref16}. These methods are generally recursive in nature. A computation scheme that computes for a given $\rho\in(0,1)$ and integer $n\geq0$ all radial polynomials $R_n^m(\rho)$ with $m=n,n-2,...,n-2\lfloor n/2\rfloor=0$ or 1, is based on the integral representation
\beq \label{e24}
R_n^m(\rho)=\frac{1}{2\pi}\,\il_0^{2\pi}\,U_n(\rho\cos\vart)\cos m\vart\,d\vart~,
\eq
where $U_n$ is the Chebyshev polynomial of the second kind and of degree $n$, see \cite{ref22}. The integrand at the right-hand side of Eq.~(\ref{e24}) is a trigonometric polynomial of degree $n+m$, and so the integration can be discretized error-free when $N>n+m$ equidistant sample points in the interval $[0,2\pi]$ are used. This then yields a DCT-based method, with all its well-known advantages in terms of speed and accuracy, to compute $R_n^m(\rho)$, $m=n,n-2,...,n-2\lfloor n/2\rfloor$.

In \cite{ref23}, the integral representation in Eq.~(\ref{e24}) is used to show the 4-terms recursion
\beq \label{e25}
R_n^{|m|}=\rho\,[R_{n-1}^{|m-1|}+R_{n-1}^{|m+1|}]-R_{n-2}^{|m|}~,
\eq
from which all radial polynomials at a particular value $\rho$ can be computed using the initialization $R_0^0\equiv1$, $R_n^{|m|}\equiv0$, $n<|m|$. This recursion is reminiscent of the identity in Eq.~(\ref{e22}), except that all coefficients in Eq.~(\ref{e25}) are $\pm\,1$, which makes Eq.~(\ref{e25}) very attractive for implementation. A more straightforward method than the one used in \cite{ref23} to show Eq.~(\ref{e25}) is to use the (generalization of the) basic recursions \cite{ref11}, (2, 25--26) in Nijboer's thesis,
\begin{eqnarray} \label{e26}
\rho\,R_n^{|m|}(\rho) & = & \frac{n+m+2}{2(n+1)}\,R_{n+1}^{|m+1|}+\frac{n-m}{2(n+1)}\, R_{n-1}^{|m+1|} \nonumber \\[3mm]
& = & \frac{n-m+2}{2(n+1)}\,R_{n+1}^{|m-1|}+\frac{n+m}{2(n+1)}\,R_{n-1}^{|m-1|}~.
\end{eqnarray}
It is finally interesting to note that the two Lukosz identities in Eq.~(\ref{e5}) follow from Eq.~(\ref{e24}), using some of the more basic properties of the Gegenbauer polynomials $C_n^{\alpha}$, of which $U_n$ is the case with $\alpha=1$.

\section{Zernike-based solution of a basic problem in wave-front sensing} \label{sec3}

In this section we consider the problem of estimating the Zernike coefficients of a wave-front function $W(\nu,\mu)$ from the Zernike coefficients of the first-order partial derivatives of $W$. At this point, we do not want to be more specific about how the latter coefficients have been obtained (analytically, semi-analytically or numerically, experimentally from matching on a set of sample points on the unit disk, etc.).

We start with the Zernike expansion of $W$,
\beq \label{e27}
W(\nu,\mu)=\sum_{m={-}\infty}^{\infty}\:\sum_{n=|m|(2)\infty}\,\alpha_n^m\,Z_n^m(\nu,\mu)~,
\eq
with
\beq \label{e28}
\osal=(\alpha_n^m)_{\footnotesize{\ba{l} m={-}\infty,...,\infty, \\ n=|m|(2)\infty\ea}}~,
\eq
an aggregate of unknown Zernike coefficients, and where $n=|m|(2)\infty$ denotes $n=|m|,|m|+2,...\,$. When considering aggregates as in Eq.~(\ref{e28}), it is sometimes convenient to set $\alpha_n^m=0$ when $n<|m|$. We assume that the Zernike coefficients of $\frac{\partial W}{\partial\nu}$ and $\frac{\partial W}{\partial\mu}$ are available, and we let
\beq \label{e29}
\overline{\osbe}_{\pm}=((\overline{\beta}_{\pm})_n^m)_{\footnotesize{\ba{l} m={-}\infty,...,\infty, \\ n=|m|(2)\infty\ea}}
\eq
be the aggregates of the Zernike coefficients of $\frac{\partial W}{\partial\nu}\pm i\,\frac{\partial W}{\partial\mu}$, so that
\beq \label{e30}
\frac{\partial W}{\partial\nu}\pm i\,\frac{\partial W}{\partial\mu}=\sum_{m={-}\infty}^{\infty}\:\sum_{n=|m|(2)\infty}\, (\overline{\beta}_{\pm})_n^m\,Z_n^m(\nu,\mu)~.
\eq
It is shown in Appendix~A from Eq.(\ref{e10}) that the analytical aggregates $\osbe_{\pm}$ of Zernike coefficients of $\frac{\partial W}{\partial\nu}\pm i\,\frac{\partial W}{\partial\mu}$ are given in terms of the aggregate $\osal$ of Zernike coefficients of $W$ by
\beq \label{e31}
\osbe_{\pm}=A_{\pm}\osal=\Bigl(2(n+1)\,\sum_{n'=n(2)\infty}\,\alpha_{n'+1}^{m\mp1} \Bigr)_{\footnotesize{\ba{l} m={-}\infty,...,\infty, \\ n=|m|(2)\infty\ea}}~.
\eq
In the matrix-vector notation just developed, we should therefore choose $\osal$ such that a best match occurs between $\overline{\osbe}_{\pm}$ of Eqs.~(\ref{e29}--\ref{e30}) and $A_{\pm}\osal$ of Eq.~(\ref{e31}).

In the space $ZC$ of aggregates $\osga=(\gamma_n^m)_{m={-}\infty,...,\infty,\,n=|m|(2)\infty}$ of Zernike coefficients, we take as inner product norm
\beq \label{e32}
\|\osga\|_{ZC}^2=(\osga,\osga)_{ZC}=\sum_{m={-}\infty}^{\infty}\:\sum_{n=|m|(2)\infty}
\, \frac{|\gamma_n^m|^2}{2(n+1)}~.
\eq
This inner product is consistent with the normalization condition
\beq \label{e33}
\il\hspace*{-6mm}\il_{\nu^2+\mu^2\leq1}\,|Z_n^m(\nu,\mu)|^2\,d\nu d\mu=\frac{1}{2(n+1)}
\eq
of the circle polynomials. Thus, we choose $\osal\in ZC$ such that
\beq \label{e34}
\|A\osal-\overline{\osbe}\|_{ZC^2}^2=\|A_+\osal-\overline{\osbe}_+\|_{ZC}^2+\| A_-\osal-\overline{\osbe}_-\|^2_{ZC}
\eq
is minimal. Here we have set in a symbolic matrix notation
\beq \label{e35}
A=\Bigl[\frac{A_+}{A_-}\Bigr]\,;~~A\osal=\Bigl[\frac{A_+\osal}{A_-\osal}\Bigr] \in ZC^2~,~~~~~~\overline{\osbe}=\Bigl[\!\ba{c} \overline{\osbe}_+ \\ \overline{\osbe}_-\ea\!\Bigr]\in ZC^2~,
\eq
and $\|\osga\|_{ZC^2}=|\osga_+\|_{ZC}^2+\|\osga_-\|_{ZC}^2$ is the inner product norm for a $\osga=[\!\ba{c} \osga_+ \\ \osga_-\ea\!]\in ZC^2$.

The least-squares $\osal$ is found by the usual linear algebra methods of generalized inverses as
\beq \label{e36}
\hat{\osal}=(A^HA)^{-1}\,A^H\,\overline{\osbe}~,
\eq
where $A^H$ is the adjoint of the operator $A$ in Eq.~(\ref{e35}), relative to the inner product $(~,~)_{ZC^2}$ in $ZC^2$. The operator $A^H$ is computed in Appendix~A as
\beq \label{e37}
A^H\,\osde=A_+^H\,\osde_+ + A_-^H\,\osde_-~,~~~~~~\osde=\Bigl[\!\ba{c} \osde_+ \\ \osde_- \ea\!\Bigr]\in ZC^2~,
\eq
with
\beq \label{e38}
A_{\pm}^H\,\osga=\Bigl(2(n+1)\,\sum_{n'=|m|(2)n}\,\gamma_{n'-1}^{m\pm1}\Bigr)_
{\footnotesize{\ba{l} m={-}\infty,...,\infty, \\ n=|m|(2)\infty\ea}}~,~~~~~~\osga\in ZC~.
\eq
The operator $A^HA$ is computed in Appendix~A as
\beq \label{e39}
A^HA\,\osga=\Bigl(4(n+1)\,\sum_{n'=|m|(2)\infty}\,B_{n\wedge n'}^m\,\gamma_{n'}^m\Bigr)_
{\footnotesize{\ba{l} m={-}\infty,...,\infty, \\ n=|m|(2)\infty\ea}}~,~~~~~~\osga\in ZC~,
\eq
where
\beq \label{e40}
n\wedge n'=\min(n,n')
\eq
and, for $n''=|m|(2)\infty$,
\beq \label{e41}
B_{n''}^m=|m|+\tfrac12\,(n''-|m|)(n''+|m|+2)~.
\eq
Thus $\hat{\osal}$ is found by solving
\beq \label{e42}
A^HA\,\osal=A^H\,\overline{\osbe}~.
\eq
It follows from Eqs.~(\ref{e37}--\ref{e38}) that
\beq \label{e43}
(A^H\,\overline{\osbe})_n^m=2(n+1)\,\sum_{n'=|m|(2)n}\,((\overline{\beta}_+)_{n'-1}^{m+1} +(\overline{\beta}_-)_{n'-1}^{m-1})~,
\eq
and so, by Eq.~(\ref{e39}), we should find the $\alpha$'s from
\beq \label{e44}
\sum_{n'=|m|(2)\infty}\,B_{n\wedge n'}^m\,\alpha_{n'}^m=\psi_n^m~,~~~~~~n=|m|(2)\infty~,
\eq
where we have set
\beq \label{e45}
\psi_n^m=\sum_{n'=|m|(2)n}\,\tfrac12\,((\overline{\beta}_+)_{n'-1}^{m+1}+ (\overline{\beta}_-)_{n'-1}^{m-1})~,
\eq
and where the $B$'s are given by Eqs.~(\ref{e40}--\ref{e41}). Note that $B_0^0=\psi_0^0$, and so Eq.~(\ref{e44}) with $m=n=0$ leaves $\alpha_0^0$ undetermined. Also note that Eq.~(\ref{e44}) allows solving $a_{n'}^m$ per separate $m$.

The linear system in Eq.~(\ref{e44}) is considered for finite $I=0,1,...\,$, and assumes the particular simple form
\beq \label{e46}
\sum_{j=0}^I\,M_{ij}\,x_j=c_i~,~~~~~~i=0,1,...,I~,
\eq
where $M$ is an $(I+1)\times(I+1)$ matrix of the form
\beq \label{e47}
M=(b_{\min(i,j)})_{i,j=0,1,...,I}~.
\eq
The right-hand side ${\bf c}=(c_i)_{i=0,1,...,I}$ of Eq.~(\ref{e46}) has the form
\beq \label{e48}
{\bf c}=L_1\,{\bf d}~,
\eq
where, see Eqs.(\ref{e44}, \ref{e45})
\beq \label{e49}
{\bf d}=(d_k)_{k=0,1,...,I}=(\varp_{|m|+2k}^m)_{k=0,1,...,I}~,
\eq
with
\beq \label{e50}
\varp_{n'}^m=\tfrac12\,(\overline{\beta}_+)_{n'-1}^{m+1}+\tfrac12\,(\overline{\beta}_-)_{n'-1}^{m-1}~,
\eq
and $L_1$ is the $(I+1)\times(I+1)$ lower triangular matrix with all entries 1 on and below the main diagonal. This $L_1$ is the inverse $L^{-1}$ of the lower triangular, bidiagonal matrix $L$ considered in Eq.~(\ref{b1}). Thus the required solution ${\bf x}=(x_j)_{j=0,1,...,I}$ is given as
\beq \label{e51n}
{\bf x}=M^{-1}L_1\,{\bf d}=(LM)^{-1}\,{\bf d}~.
\eq
In Appendix~B it is shown that
\beq \label{e52n}
(LM)^{-1}=\left[\ba{ccccc}
a_0 & -a_1 & 0 & & 0 \\[1mm]
0 & a_1 & -a_2 \\[1mm]
& & & & \\[1mm]
& & & & \\[1mm]
& & & a_{I-1} & -a_I \\[1mm]
0 & & & 0 & a_I
\ea\right]
\eq
(upper triangular, bidiagonal $(I+1)\times(I+1)$ matrix with $a_0,a_1,...,a_I$ on the main diagonal and $-a_1,{-}a_2,...,{-}a_I$ on the first upper co-diagonal), where
\beq \label{e53n}
a_j=\frac{1}{b_j-b_{j-1}}~,~~~~~~j=0,1,...,I~,
\eq
with $b_j$ from Eq.~(\ref{e47}) and where we have set $b_{-1}=0$. Thus we get
\beq \label{e54n}
x_j=a_jd_j-a_{j+1}d_{j+1}\,,~~j=0,1,...,I-1~;~~~~~~x_I=a_Id_I~.
\eq

In the present case, we have
\beq \label{e55n}
b_j=B_{|m|+2j}^m=|m|+2j(|m|+j+1)~,~~~~~~j=0,1,...,I~,
\eq
and so we get
\beq \label{e56n}
a_0=\frac{1}{|m|}~;~~~~~~a_j=\frac{1}{2(|m|+2j)}\,,~~j=1,2,...,I~.
\eq
For the case $m=0$, we have $b_0=B_0^0=0$, and one should delete the first row and column of the $M$-matrix in Eq.~(\ref{e47}) and solve for $x_1,x_2,...,x_I$.

With $d_j$ given in Eq.~(\ref{e49}) and $a_j$ given in Eq.~(\ref{e56n}), we can then write the solution of Eq.~(\ref{e54n}) in terms of the $\hat{\alpha}_n^m$ as
\beq \label{e57n}
\hat{\alpha}_n^m=C_n^m\,\varp_n^m-C_{n+2}^m\,\varp_{n+2}^m~,~~~~~~n=|m|(2)(|m|+I-2)~,
\eq
\beq \label{e58n}
\hat{\alpha}_{|m|+2I}^m=C_{|m|+2I}^m\,\varp_{|m|+2I}^m~,
\eq
where $\varp_n^m$ are given by Eq.~(\ref{e50}) and
\beq \label{e59n}
C_{|m|}^m=\frac{1}{|m|}\,,~~n=|m|~;~~~~~~C_n^m=\frac{1}{2n}\,,~~n=(|m|+2)(2)(|m|+2I)~.
\eq
In Eq.~(\ref{e57n}, we consider $n=2(2)(I-2)$ in the case that $m=0$.

It is important to note that the finitization to linear systems of order $(I+1)\times(I+1)$ has virtually no influence on the computed $\hat{\alpha}_n^m$, except for the validity range $n=|m|(2)(|m|+2I-2)$.

It is relatively straightforward to check that Eqs.(\ref{e57n}, \ref{e59n}) yield $\alpha_n^m=\delta_{mm_1}\;\delta_{nn_1}$ when $W=Z_{n_1}^{m_1}$ and the partial derivative data $\overline{\osbe}_{\pm}$ are perfect (integer $n_1,m_1$ with $n_1-|m_1|$ even and non-negative, and $0<n_1\leq |m_1|+2I-2$).
\section{Zernike expansion of the Laplacian and the inverse Laplacian of circle polynomials} \label{sec4}

Higher-order partial derivatives of wave-front functions occur in various places in the optics literature on wave-front reconstruction and ophthalmics. In \cite{ref24}, wave-front reconstruction from defocused images in an astronomical setting is considered, and for this conservation of the intensity flux requires studying Jacobians of transformations that involve the second-order partial derivatives of the aberration. In \cite{ref8}, a detailed study is made of the action of the Laplacian on the linear space ${\cal Z}_{N+2}$ spanned by the circle polynomials with degree not exceeding $N+2$ as a mapping into ${\cal Z}_N$ in connection with the Transport-of-Intensity equation. In particular, one is interested in solving the Neumann problem of finding $\varp\in {\cal Z}_{N+2}$ from $-\Delta\varp=f\in {\cal Z}_N$ with $\partial_{{\bf n}}\varp=\psi$ on the boundary of the disk. Such a problem is also considered, with boundary function $\psi=0$, in \cite{ref25}. Finally, in \cite{ref15a}, one is in principle interested in the partial derivatives of the circle polynomials of all orders in order to study the effect of decentration of the optical system. Here one aims at obtaining a Zernike expansion of a displaced wave-front function $W(\nu+\nu_0,\mu+\mu_0)$ from such an expansion of $W(\nu,\mu)$ by Taylor expansion. For the latter problem, we may refer also to \cite{ref26}, where a completely analytic solution is given for the problem of expanding any shifted-and-scaled circle polynomial into the set of circle polynomials orthogonal on the original reference disk (also see \cite{ref18} for the pure-scaling case).

In this section, we concentrate on the Laplacians of the circle polynomials, and we show that for integer $n$ and $m$ such that $n-|m|$ is even and non-negative
\beq \label{e51}
\Bigl(\frac{\partial^2}{\partial\nu^2}+\frac{\partial^2}{\partial\mu^2}\Bigr)\,Z_n^m=4\, \sum_{t=0}^{\frac12(n-|m|)-1}\,(n-1-2t)(n-t)(t+1)\,Z_{n-2-2t}^m~.
\eq
From this formula, many of the observations made in \cite{ref8} are verified instantly. Sinilar, but somewhat more complicated, series expressions can be shown to hold for the three second-order partial derivatives of $Z_n^m$ separately.

To show Eq.~(\ref{e51}), we observe that
\beq \label{e52}
\frac{\partial^2}{\partial\nu^2}+\frac{\partial^2}{\partial\mu^2}= \Bigl(\frac{\partial}{\partial\nu}+i\,\frac{\partial}{\partial\mu}\Bigr)
\Bigl(\frac{\partial}{\partial\nu}-i\,\frac{\partial}{\partial\mu}\Bigr)~,
\eq
and so, we get from Eq.~(\ref{e10}) with summation upper limits $\frac12\,(n-1-|m\pm1|)$ that
\begin{eqnarray} \label{e53}
& \mbox{} & \Bigl(\frac{\partial^2}{\partial\nu^2}+\frac{\partial^2}{\partial\mu^2}\Bigr) \,Z_n^m = 2\,\sum_{l=0}^{\frac{n-1-|m-1|}{2}}\,(n-2l)\Bigl(\frac{\partial}{\partial\nu}+i\,\frac{\partial}{\partial\mu} \Bigr)\,Z_{n-1-2l}^{m-1} \nonumber \\[3.5mm]
& & =~2\,\sum_{l=0}^{\frac{n-1-|m-1|}{2}}\:\sum_{k=1}^{\frac{n-1-2l-1-|m-1+1|}{2}}\, 2(n-1-2l-2k)\,Z_{n-1-2l-1-2k}^{m-1+1} \nonumber \\[3.5mm]
& & =~4\,\sum_{l=0}^{\frac{n-1-|m-1|}{2}}\:\sum_{k=0}^{\frac{n-2l-2-|m|}{2}}\, (n-2l)(n-1-2l-2k)\, Z_{n-2-2l-2k}^m~.
\end{eqnarray}
We must now carefully rearrange the double series on the last line of Eq.~(\ref{e53}). First, we have that
\beq \label{e54}
\frac{n-1-|m-1|}{2}=\left\{\ba{llll}
\tfrac12\,(n-|m|) & \!\!, & ~~~m\geq1 & \!\!, \\[2mm]
\tfrac12\,(n-|m|)-1 & \!\!, & ~~~m\leq0 & \!\!.
\ea\right.
\eq
In the case that $m\geq1$ and $l=\frac12\,(n-|m|)$, we have that the summation over $k$ in the last line of Eq.~(\ref{e53}) is empty, and so we can delete the term with $l=\frac12\,(n-|m|)$. Hence, in all cases
\beq \label{e55}
\Bigl(\frac{\partial^2}{\partial\nu^2}+\frac{\partial^2}{\partial\mu^2}\Bigr)\,Z_n^m=4 \,\sum_{l=0}^{p-1}\:\sum_{k=0}^{p-1-l}\,(n-2l)(n-1-2l-2k)\,Z_{n-2-2l-2k}^m~,
\eq
where we have set $p=\frac12\,(n-|m|)$. The summation in Eq.~(\ref{e55}) can be written as
\begin{eqnarray} \label{e56}
& \mbox{} & 4\,\sum_{t=0}^{p-1}\:\sum_{l,k\geq0\,;\,l+k=t}\,(n-2l)(n-1-2t)\,Z_{n-2-2t}^m \nonumber \\[3.5mm]
& & =~4\,\sum_{t=0}^{p-1}\,(n-1-2t)\,Z_{n-2-2t}^m\,S_{nt}~,
\end{eqnarray}
where
\beq \label{e57}
S_{nt}=\sum_{l,k\geq0\,;\,l+k=t}\,(n-2l)=\sum_{l=0}^t\,(n-2l)=(n-t)(t+1)~,
\eq
and we arrive at the result of Eq.~(\ref{e51}).
\begin{center}
\os{Table I}: Equation~(\ref{e51}) for all circle polynomials $Z_n^m$ with $m\geq0$, $n\leq6$. \\[8mm]
$\ba{lcl}
\Delta Z_0^0 & = & \Delta 1=0~, \\[2mm]
\Delta Z_2^0 & = & \Delta(2(\nu^2+\mu^2)-1)=8=8Z_0^0~, \\[2mm]
\Delta Z_4^0 & = & \Delta(6(\nu^2+\mu^2)^2-6(\nu^2+\mu^2)+1)=96(\nu^2+\mu^2)-24~= \\[2mm]
& & 48Z_2^0+24Z_0^0~, \\[2mm]
\Delta Z_6^0 & = & \Delta(20(\nu^2+\mu^2)^3-30(\nu^2+\mu^2)^2+12(\nu^2+\mu^2)-1)~= \\[2mm]
& & 720(\nu^2+\mu^2)^2-480(\nu^2+\mu^2)+48=120Z_4^0+120Z_2^0+48Z_0^0~, \\[6mm]
\Delta Z_1^1 & = & \Delta(\nu+i\mu)=0~, \\[2mm]
\Delta Z_3^1 & = & \Delta\,[(3(\nu^2+\mu^2)-2)(\nu+i\mu)]=24(\nu+i\mu)=24Z_1^1~, \\[2mm]
\Delta Z_5^1 & = & \Delta\,[(10(\nu^2+\mu^2)^2-12(\nu^2+\mu^2)+3)(\nu+i\mu)]~= \\[2mm]
& & (240(\nu^2+\mu^2)-96)(\nu+i\mu)=80Z_3^1+64Z_1^1~, \\[6mm]
\Delta Z_2^2 & = & \Delta(\nu+i\mu)^2=0~, \\[2mm]
\Delta Z_4^2 & = & \Delta\,[(4(\nu^2+\mu^2)-3)(\nu+i\mu)^2]=48(\nu+i\mu)^2=48Z_2^2~, \\[2mm]
\Delta Z_6^2 & = & \Delta\,[(15(\nu^2+\mu^2)^2-20(\nu^2+\mu^2)+6)(\nu+i\mu)^2]~= \\[2mm]
& & (480(\nu^2+\mu^2)-240)(\nu+i\mu)^2=120Z_4^2+120Z_2^2~, \\[6mm]
\Delta Z_3^3 & = & \Delta(\nu+i\mu)^3=0~, \\[2mm]
\Delta Z_5^3 & = & \Delta\,[(5(\nu^2+\mu^2)-4)(\nu+i\mu)^3]=80(\nu+i\mu)^3=80Z_3^3~, \\[6mm]
\Delta Z_4^4 & = & \Delta(\nu+i\mu)^4=0~, \\[2mm]
\Delta Z_6^4 & = & \Delta\,[(6(\nu^2+\mu^2)-5)(\nu+i\mu)^4]=120(\nu+i\mu)^4=120Z_4^4~, \\[6mm]
\Delta Z_6^6 & = & \Delta(\nu+i\mu)^6=0~.
\ea$
\end{center}
\mbox{} \\ \\
The result of Eq.~(\ref{e51}) can be checked in particular cases by writing down the $Z_n^m(\rho,\vart)=R_n^{|m|}(\rho)\,\exp[im\vart]$ in Cartesian coordinates and performing the differentiations implied by $\Delta=\frac{\partial^2}{\partial\nu^2}+\frac{\partial^2}{\partial\mu^2}$. For instance, when $m=2$, $n=6$, this leads to
\newpage
\begin{eqnarray} \label{e58}
& \mbox{} & \Bigl(\frac{\partial^2}{\partial\nu^2}+\frac{\partial^2}{\partial\mu^2}\Bigr)\,Z_6^2(\nu,\mu) \nonumber \\[3mm]
& & =~\Bigl(\frac{\partial^2}{\partial\nu^2}+\frac{\partial^2}{\partial\mu^2}\Bigr)((15(\nu^2+\mu^2)^2-20(\nu^2+\mu^2)+6)(\nu+i\mu)^2) \nonumber \\[3mm]
& & =~(480(\nu^2+\mu^2)-240)(\nu+i\mu)^2=120Z_4^2+120Z_2^2~,
\end{eqnarray}
in which the second step requires a lengthy computation. In Table~I, we illustrate Eq.~(\ref{e51}) for all circle polynomials $Z_n^m$ with $m\geq0$, $n\leq6$.

With the notation introduced in Eq.~(\ref{e9}), we can write Eq.~(\ref{e51}) concisely as
\beq\ \label{e59}
\Delta Z_n^m=\sum_{s=|m|(2)(n-2)}\,(s+1)(n+s+2)(n-s)\,Z_s^m~.
\eq

We next consider, as in \cite{ref8}, for a given integer $N\geq0$, the Neumann problem
\beq \label{e60}
-\Delta\varp=f~,~~~~~~\partial_{{\bf n}}\varp=\psi~,
\eq
where $f$ belongs to the linear space $Z_N$ spanned by all circle polynomials of degree $\leq\,N$, and $\psi=\psi(\vart)$ is a periodic function of degree $\leq\,N+2$ (and thus a linear combination of $\exp[im\vart]$, integer $m$, $|m|\leq N+2$).

We let ${\cal Z}_N^m$ and ${\cal Z}_{N+2}^m$, for integer $m$ with $|m|\leq N$, be the spaces spanned by the circle polynomials of azimuthal order $m$ and of degree $\leq\,N$ and $N+2$, respectively. It is seen from Eq.~(\ref{e59}) that $\Delta$ maps ${\cal Z}_{N+2}^m$ into ${\cal Z}_N^m$ and that $\Delta Z_{|m|}^m=0$. Furthermore, the matrix $B^m$ of $\Delta$, when choosing in ${\cal Z}_N^m$ and ${\cal Z}_{N+2}^m$ the orthogonal basis of circle polynomials of azimuthal order $m$ and degree $\leq\,N$ and $N+2$, respectively, is lower triangular. The matrix elements $B_{ns}^m$ are given by
\beq \label{e61}
B_{ns}^m=\left\{\ba{llll}
(s+1)(n+s+2)(n-s) & \!\!, & ~~~s=|m|(2)(n-2) & \!\!, \\[2mm]
0 & \!\!, & ~~~s=n(2)\,N_m & \!\!,
\ea\right.
\eq
where
\beq \label{e62}
N_m=|m|+2\lfloor\tfrac12\,(N-|m|)\rfloor
\eq
so that $N_m=N$ or $N-1$ according as $N$ and $|m|$ have same or opposite parity. Observing that $B_{n,n-2}\neq0$, it follows that the functions $\Delta Z_n^m$, $n=(|m|+2)(2)(N_m+2)$ are independent, and so $\Delta$ maps ${\cal Z}_{N+2}^m$ onto ${\cal Z}_N^m$.

For solving $-\Delta\varp=f$, see Eq.~(\ref{e60}), we develop $f$ as
\beq \label{e63}
f=\sum_{m={-}N}^N\:\sum_{n'=|m|(2)N_m}\,\alpha_{n'}^m\,Z_{n'}^m~.
\eq
\begin{center}
\os{Table II}: Equation~(\ref{e69}) for all circle polynomials $Z_{n'}^m$ \\ with $m\geq0$ and $n'\leq6$. \\[8mm]
$\ba{lcl}
Z_0^0 & = & \Delta\,[\frac18\,Z_2^0] \\[2mm]
Z_2^0 & = & \Delta\,[\frac{-1}{16}\,Z_2^0+\frac{1}{48}\,Z_4^0] \\[2mm]
Z_4^0 & = & \Delta\,[\frac{1}{80}\,Z_2^0-\frac{1}{48}\,Z_4^0+\frac{1}{120}\,Z_6^0] \\[2mm]
Z_6^0 & = & \Delta\,[\frac{1}{168}\,Z_4^0-\frac{1}{96}\,Z_6^0+\frac{1}{224}\,Z_8^0] \\[6mm]
Z_1^1 & = & \Delta\,[\frac{1}{24}\,Z_3^1] \\[2mm]
Z_3^1 & = & \Delta\,[\frac{-1}{30}\,Z_3^1+\frac{1}{80}\,Z_5^1] \\[2mm]
Z_5^1 & = & \Delta\,[\frac{1}{120}\,Z_3^1-\frac{1}{70}\,Z_5^1+\frac{1}{168}\,Z_7^1] \\[6mm]
Z_2^2 & = & \Delta\,[\frac{1}{48}\,Z_4^2] \\[2mm]
Z_4^2 & = & \Delta\,[\frac{-1}{48}\,Z_4^2+\frac{1}{120}\,Z_6^2] \\[2mm]
Z_6^2 & = & \Delta\,[\frac{1}{168}\,Z_4^2-\frac{1}{96}\,Z_6^2+\frac{1}{224}\,Z_8^2] \\[6mm]
Z_3^3 & = & \Delta\,[\frac{1}{80}\,Z_5^3] \\[2mm]
Z_5^3 & = & \Delta\,[\frac{-1}{70}\,Z_5^3+\frac{1}{168}\,Z_7^3] \\[6mm]
Z_4^4 & = & \Delta\,[\frac{1}{120}\,Z_6^4] \\[2mm]
Z_6^4 & = & \Delta\,[\frac{-1}{96}\,Z_6^4+\frac{1}{224}\,Z_8^4] \\[6mm]
Z_5^5 & = & \Delta\,[\frac{1}{168}\,Z_7^5] \\[6mm]
Z_6^6 & = & \Delta\,[\frac{1}{224}\,Z_8^6]
\ea$
\end{center}
\mbox{} \\ \\
It is seen that the basic problem is to solve $\beta_n^m$ for a given $m$, $|m|\leq N$, and a given $n'=|m|(2)N_m$ from the linear equations
\beq \label{e64}
\Delta\Bigl(\sum_{n=(|m|+2)(2)(N_m+2)}\,\beta_n^m\,Z_n^m\Bigr)=Z_{n'}^m~.
\eq
In terms of the matrix elements $B_{ns}^m$ in Eq.~(\ref{e61}), the Eq.~(\ref{e64}) can be written as
\beq \label{e65}
\sum_{n=s(2)N_m}\,B_{n+2,s}^m\,\beta_{n+2}^m=\delta_{sn}~,~~~~~~s=|m|(s)N_m~.
\eq
By the properties of $B^m$ as a triangular matrix, we have
\beq \label{e66}
B_{s+2,s}^m\neq0=B_{n+2,s}^m~,~~~~~~n=|m|(2)s~,
\eq
for $s=|m|(2)N_m$, and so the $\beta_n^m$ can be solved by backward elimination. However, as is shown in Appendix~B, the linear system in Eq.~(\ref{e65}) can be solved explicitly, with the result
\beq \label{e67}
Z_{n'}^m=\Delta\,\Bigl[\frac{1}{4(n'+2)(n'+1)}\,Z_{n'+2}^m-\frac{1}{2n'(n'+2)}\,Z_{n'}^m+\frac{1}{4n'(n'+1)}\,Z_{n'-2}^m\Bigr]~,
\eq
where we recall that $\Delta Z_n^m=0$ when $n \leq |m|$. The result of Eq.~(\ref{e67}) is illustrated in Table~II.

Having solved the problem in Eq.~(\ref{e64}), and therefore, by linear combination with $f$ as in Eq.~(\ref{e63}), the problem $-\Delta\varp=f\in {\cal Z}_N$, it remains to satisfy the boundary condition $\partial_{{\bf n}}\varp=\psi$, see Eq.~(\ref{e60}). This problem can be solved in the same manner as this is done in \cite{ref8}, pp.~1936--37, where a linear combination of $Z_{|m|}^m$, $m={-}N,...,N\,$, is used to satisfy this boundary condition. To this end, it is useful to note that
\beq \label{e69}
\partial_{{\bf n}}[Z_n^m(\rho,\vart)]=(R_n^{|m|})'(1)\,\exp[im\vart]=\tfrac12\,(n(n+2)-m^2)\,\exp[im\vart]~,
\eq
so that $\partial_{{\bf n}}[Z_m^m(\rho,\vart)]=|m|\,\exp[im\vart]$. Thus, the solution $\varp_0$ of $-\Delta\varp_0=f$, not comprising any $Z_{|m|}^m$, has computable Fourier coefficients of its normal derivative $\partial_{{\bf n}}\varp_0=\psi_0$, and the coefficients $c_m$ of $Z_{|m|}^m$ in the full $\varp$ should be chosen such that $|m|\,c_m$ equals the $m^{{\rm th}}$ Fourier coefficient of $\psi-\psi_0$.

\section{Conclusions} \label{sec5}

We have given a review of the results concerning the first-order Cartesian derivatives of the Zernike circle polynomials. By choosing the version of the circle polynomials with exponential azimuthal dependence and by proper combination of the two first-order partial derivatives, the results have been brought into a concise form. This form allows a convenient formulation and solution of a basic problem in wave-front sensing in which the Zernike coefficients of the wave-front function are to be estimated from the Zernike coefficients of the first-order Cartesian derivatives. It has been shown that the matrix inversion required for solving the ensuing least-squares problem can be done analytically. The preferred version of the circle polynomials together with proper combination of the first-order derivatives also leads to a concise result for the Laplacians of the circle polynomials. This concise result has been used to find an explicit formula for the inverse Laplacian of any circle polynomial. This yields a concise solution of the Neumann problem, that occurs when studying the Transport-of-Intensity equation on spaces of circle polynomials with radial degree not exceeding a fixed number.

\setcounter{equation}{0}
\renewcommand{\theequation}{A\arabic{equation}}
\section*{Appendix A: Proofs of the properties of $A_{\pm}$ in Section~\ref{sec3}} \label{appA}

In this appendix, we prove the results on the operators $A_{\pm}$ and $A$ as used in Section~\ref{sec3}. We start with the proof of Eq.~(\ref{e31}) that expresses the aggregate of Zernike expansion coefficients $\osbe_{\pm}$ of $\frac{\partial W}{\partial\nu}\pm i\,\frac{\partial W}{\partial\mu}$ in terms of the corresponding aggregate $\osal$ of $W$, see Eq.~(\ref{e27}). With integer $m_1$, $n_1$, such that $n_1-|m_1|$ is even and non-negative, we shall verify Eq.~(\ref{e31}) directly for the case that $W=Z_{n_1}^{m_1}$ so that $\alpha_n^m=\delta_{m_1m}\,\delta_{n_1n}$. The identity to be verified is
\beq \label{a1}
\frac{\partial W}{\partial\nu}\pm i\,\frac{\partial W}{\partial\mu}=2\,\sum_{m={-}\infty}^{\infty}\:\sum_{n=|m|(2)\infty}\,(n+1)\Bigl(\sum_{n'=n(2)\infty}\,\alpha_{n'+1}^{m\mp1}\Bigr)\,Z_n^m~.
\eq
With $\alpha_n^m=\delta_{m_1m}\,\delta_{n_1n}$, it is seen that in the series over $m$ in Eq.~(\ref{a1}) only the term $m$ with $m\mp1=m_1$ is non-vanishing, and so we should only consider
\beq \label{a2}
2\,\sum_{n=|m_1\pm1|(2)\infty}\,\Bigl(\sum_{n'=n(2)\infty}\,(n+1)\,\alpha_{n'+1}^{m_1}\Bigr)\,Z_n^{m_1\pm1}~.
\eq
With $n=|m_1\pm1|(2)\infty$, we have that
\beq \label{a3}
\sum_{n'=n(2)\infty}\,\alpha_{n'+1}^{m_1}=\left\{\ba{llll}
1 & \!\!, & ~~~n+1\leq n_1 & \!\!, \\[3mm]
0 & \!\!, & ~~~{\rm otherwise} & \!\!.
\ea\right.
\eq
Hence, the expression in Eq.~(\ref{a2}) equals
\begin{eqnarray} \label{a4}
& \mbox{} & 2\,\sum_{n=|m_1\pm1|(2)(n_1-1)}\,(n+1)\,Z_n^{m_1\pm1}=2\,\sum_{n=(|m_1\pm1|+1)(2)n_1}\,n\,Z^{m_1\pm1} \nonumber \\[3.5mm]
& & =~2\,\sum_{l=0}^{\frac12(n_1-1-|m_1\pm1|)}\,(n_1-2l)\,Z_{n_1-1-2l}^{m_1\pm1}=\Bigl(\frac{\partial}{\partial\nu}\pm i\,\frac{\partial}{\partial\mu}\Bigr) \,Z_{n_1}^{m_1}
\end{eqnarray}
according to Eq.~(\ref{e10}) with summation of non-zero terms only. This shows Eq.~(\ref{e31}) for the case that $W=Z_{n_1}^{m_1}$, and the general case follows from this by linear superposition of terms $Z_{n_1}^{m_1}$ in $W$ with integer $m_1$ and $n_1$ such that $n_1-|m_1|$ is even and non-negative.

We shall now compute $A^H$ and $A^HA$, as required in Eq.~(\ref{e36}) for the least squares $\osal$. With $\osga,\osde\in ZC^2$, written as
\beq \label{a5}
\osga=\Bigl[\ba{c} \osga_+ \\ \osga_- \ea\Bigr]~,~~~~~~\osde=\Bigl[\ba{c} \osde_+ \\ \osde_- \ea\Bigr]~,
\eq
where $\osga_{\pm},\osde_{\pm}\in ZC$ and with the notation of Eq.~(\ref{e35}), we have
\begin{eqnarray} \label{a6}
& \mbox{} & (A\osga,\osde)_{ZC^2}=(A_+\osga_+,\osde_+)_{ZC}+(A_-\osga_-,\osde_-)_{ZC} \nonumber \\[3mm]
& & =~(\osga_+,A_+^H\osde_+)_{ZC}+(\osga_-,A_-^H\osde_-)_{ZC}=(\osga,A^H\osde)_{ZC^2}~,
\end{eqnarray}
where $A_{\pm}^H$ are the adjoints of $A_{\pm}$ and
\beq \label{a7}
A^H\osde=A_+^H\osde_++A_-^H\osde_-~.
\eq
We therefore need to determine $A_{\pm}^H$.

As to the $+$-case, we have for $\osal\in ZC$ by Eq.~(\ref{e31})
\beq \label{a8}
A_+\osal=\Bigl(2(n+1)\,\sum_{n'=n(2)\infty}\,\alpha_{n'+1}^{m-1}\Bigr)_{\footnotesize{\ba{l} m={-}\infty,...,\infty, \\ n=|m|(2)\infty\ea}}
\eq
Hence, with the definition of the inner product in $ZC$, see Eq.~(\ref{e32}), we have for $\osal,\osbe\in ZC$ that
\begin{eqnarray} \label{a9}
(A_+\osal,\osbe)_{ZC} & = & \sum_{m={-}\infty}^{\infty}\:\sum_{n=|m|(2)\infty}\:\frac{(A^+\osal)_n^m\,(\beta_n^m)^{\ast}}{2(n+1)} \nonumber \\[3.5mm]
& = & \sum_{m={-}\infty}^{\infty}\,\Bigl(\sum_{n=|m|(2)\infty}\,\Bigl(\sum_{n'=n(2)\infty}\,\alpha_{n'+1}^{m-1}\Bigr)\,(\beta_n^m)^{\ast}\Bigr)
\nonumber \\[3.5mm]
& = & \sum_{m={-}\infty}^{\infty}\,\Bigl(\sum_{n'=|m|(2)\infty}\:\sum_{n=|m|(2)n'}\,\alpha_{n'+1}^{m-1}\,(\beta_n^m)^{\ast}\Bigr)
\nonumber \\[3.5mm]
& = & \sum_{m={-}\infty}^{\infty}\,\Bigl(\sum_{n'=(|m+1|+1)(2)\infty}\:\sum_{n=|m+1|(2)(n'-1)}\,\alpha_{n'}^m\,(\beta_n^{m+1})^{\ast}\Bigr)~. \nonumber \\
\mbox{}
\end{eqnarray}
Now $|m+1|=m+1$ when $m=0,1,...$ and $|m+1|=|m|-1$ when $m={-}1,{-}2,...\,$, and so
\begin{eqnarray} \label{a10}
(A_+\osal,\osbe)_{ZC} & = & \sum_{m=0}^{\infty}\,\Bigl(\sum_{n'=(m+2)(2)\infty}\:\sum_{n=(m+1)(2)(n'-1)}\,\alpha_{n'}^m\,(\beta_n^{m+1})^{\ast}\Bigr) \nonumber \\[3.5mm]
& & +~\sum_{m={-}\infty}^{-1}\,\Bigl(\sum_{n'=|m|(2)\infty}\:\sum_{n=(|m|-1)(2)(n'-1)}\,\alpha_{n'}^m\,(\beta_n^{m+1})^{\ast}\Bigr) \nonumber \\[3.5mm]
& = & \sum_{m=0}^{\infty}\,\Bigl(\sum_{n'=(m+2)(2)\infty}\:\sum_{n=(m+2)(2)n'}\,\alpha_{n'}^m\,(\beta_{n-1}^{m+1})^{\ast}\Bigr) \nonumber \\[3.5mm]
& & +~\sum_{m={-}\infty}^{-1}\,\Bigl(\sum_{n'=|m|(2)\infty}\:\sum_{n=|m|(2)n'}\,\alpha_{n'}^m\,(\beta_{n-1}^{m+1})^{\ast}\Bigr)~.
\end{eqnarray}
The terms $\beta_{n-1}^{m+1}$ with $n=m$ in the first triple series in the last member of Eq.~(\ref{a10}) vanish, and so we can extend the $n$-summation range in this triple series to $m(2)n'$. Doing so, and subsequently extending the $n'$-summation range to $m(2)\infty$ in this same triple series, we get
\begin{eqnarray} \label{a11}
(A_+\osal,\osbe)_{ZC} & = & \sum_{m=0}^{\infty}\,\Bigl(\sum_{n'=m(2)\infty}\:\sum_{n=m(2)n'}\,\alpha_{n'}^m\,(\beta_{n-1}^{m+1})^{\ast}\Bigr) \nonumber \\[3.5mm]
& & +~\sum_{m={-}\infty}^{-1}\,\Bigl(\sum_{n'=|m|(2)\infty}\:\sum_{n=|m|(2)n'}\,\alpha_{n'}^m\,(\beta_{n-1}^{m+1})^{\ast}\Bigr) \nonumber \\[3.5mm]
& = & \sum_{m={-}\infty}^{\infty}\,\Bigl(\sum_{n'=|m|(2)\infty}\:\sum_{n=|m|(2)n'}\,\alpha_{n'}^m\,(\beta_{n-1}^{m+1})^{\ast}\Bigr)~.
\end{eqnarray}
Next, interchanging the summation indices $n'$ and $n$ in the last line of Eq.~(\ref{a11}), and throwing in a factor $2(n+1)/2(n+1)$, we get
\begin{eqnarray} \label{a12}
(A_+\osal,\osbe)_{ZC} & = & \sum_{m={-}\infty}^{\infty}\,\Bigl(\sum_{n=|m|(2)\infty}\:\frac{\alpha_n^m(2(n+1)\,\dsum_{n'=|m|(2)n}\,\beta_{n+1}^{m+1})^{\ast}} {2(n+1)}\Bigr) \nonumber \\[3.5mm]
& = & (\osal,A_+^H\osbe)_{ZC}~,
\end{eqnarray}
where
\beq \label{a13}
A_+^H\osbe=\Bigl(2(n+1)\,\sum_{n'=|m|(2)n}\,\beta_{n'-1}^{m+1}\Bigr)_{\footnotesize{\ba{l} m={-}\infty,...,\infty, \\ n=|m|(2)\infty\ea}}~.
\eq

In an entirely similar fashion, we compute
\beq \label{a14}
(A_-\osal,\osbe)_{ZC}=(\osal,A_-^H\osbe)_{ZC}~,
\eq
where
\beq \label{a15}
A_-^H\osbe=\Bigl(2(n+1)\,\sum_{n'=|m|(2)\infty}\,\beta_{n'-1}^{m-1}\Bigr)_{\footnotesize{\ba{l} m={-}\infty,...,\infty, \\ n=|m|(2)\infty\ea}}~.
\eq

We finally compute $A^HA=A_+^HA_++A_-^HA$. Thus for $\osga\in ZC$, we have from Eq.~(\ref{e38}) that
\beq \label{a16}
(A_+^HA_+\osga)_n^m=2(n+1)\,\sum_{n'=|m|(2)n}\,(A_+\osga)_{n'-1}^{m+1}
\eq
when $m$, $n$ are integer and $n=|m|(2)\infty$. Now for $n'-1=|m+1|(2)\infty$, we have from Eq.~(\ref{e31}) that
\beq \label{a17}
(A_+\osga)_{n'-1}^{m+1}=2(n'+1-1)\,\sum_{n''=(n'-1)(2)\infty}\,\gamma_{n''+1}^{(m+1)-1}=2n'\,\sum_{n''=n'(2)\infty}\,\gamma_{n''}^m~.
\eq
Hence
\beq \label{a18}
(A_+^HA_+\osga)_n^m=2(n+1)\,\sum_{\footnotesize{\ba{l} n'=|m|(2)n, \\ n'-1\geq|m+1|\ea}}\, 2n'\,\sum_{n''=n'(2)\infty}\,\gamma_{n''}^m~.
\eq
In a similar fashion
\beq \label{a19}
(A_-^HA_-\osga)_n^m=2(n+1)\,\sum_{\footnotesize{\ba{l} n'=|m|(2)n, \\ n'-1\geq|m-1|\ea}}\,2n'\, \sum_{n''=n'(2)\infty}\,\gamma_{n''}^m~.
\eq
As to the conditions $n'-1\geq|m+1|$, $n'-1\geq|m-1|$ that appear in the summations in Eqs.~(\ref{a18}--\ref{a19}), we note that
\beq \label{a20}
|m+1|=\left\{\ba{lll}
|m|+1 & \!\!, & ~~~m\geq0 \\[2mm]
|m|-1 & \!\!, & ~~~m<0 \ea\right.~,~~~~~~|m-1|=\left\{\ba{llll}
|m|-1 & \!\!, & ~~~m>0 \\[2mm]
|m|+1 & \!\!, & ~~~m\leq0 & \!\!. \ea\right.
\eq
Hence, we have $n'=(|m|+2)(2)n$ when $m\geq0$ and $n'=|m|(2)n$ when $m<0$ in Eq.~(\ref{a18}), and $n'=|m|(2)n$ when $m>0$ and $n'=(|m|+2)(2)n$ when $m\leq0$ in Eq.~(\ref{a19}). Therefore,
\begin{eqnarray} \label{a21}
(A^HA\osga)_n^m & = & (A_+^HA_+\osga)_n^m+(A_-^HA_-\osga)_n^m \nonumber \\[3mm]
& = & 4(n+1)\,\sum_{n'=|m|(2)n}\:\sum_{n''=n'(2)\infty}\,n'\,\eps_{n'-|m|}\,\gamma_{n''}^m~,
\end{eqnarray}
where $\eps_k=1$ when $k=0$ and $\eps_k=2$ when $k=2,4,...$ (Neumann's symbol). The formula in Eq.~(\ref{a21}) is also valid for $m=0$, for $n'\,\eps_{n'-|m|}\,\gamma_{n''}^m$ vanishes when $n'=|m|=0$.

We rearrange the result of Eq.~(\ref{a21}) further as
\beq \label{a22}
(A^HA\osga)_n^m=4(n+1)\,\sum_{n''=|m|(2)\infty}\,\gamma_{n''}^m\,\sum_{n'=|m|(2)(n''\wedge n)}\, n'\,\eps_{n'-|m|}~,
\eq
where $k\wedge l$ is short-hand notation for $\min(k,l)$. Finally, for $n'''=|m|(2)\infty$, we have
\beq \label{a23}
\sum_{n'=|m|(2)n'''}\,n'\,\eps_{n'-|m|}=|m|+\tfrac12\,(n'''-|m|)(n'''+|m|+2)~,
\eq
and the result of Eq.~(\ref{e39}) follows from Eqs.~(\ref{a22}--\ref{a23}) upon changing $n''$ into $n'$ in $\gamma_{n''}^m$ and $n'$ into $n'''$ in $\sum_{n'}$ in Eq.~(\ref{a22}) and using Eq.~(\ref{a23}) with $n'''=n'\wedge n=n\wedge n'$.

\setcounter{equation}{0}
\renewcommand{\theequation}{B\arabic{equation}}
\section*{Appendix B: Inversion of some special matrices} \label{appB}

We shall first show that $(LM)^{-1}$ is given by Eq.~(\ref{e52n}), where $L$ is the $(I+1)\times(I+1)$ lower triangular matrix
\beq \label{b1}
L=\left[\ba{ccccc}
1 & 0 & 0 & & 0 \\[1mm]
-1 & 1 & 0 & & 0 \\[1mm]
& & & & \\[1mm]
& & & & \\[1mm]
0 & & & -1 & 1
\ea\right]
\eq
(bidiagonal matrix, with $1$'s on the main diagonal and $-1$'s on the first lower co-diagonal), and
\beq \label{b2n}
M=(b_{\min(i,j)})_{i,j=0,1,...,I}=\left[\ba{ccccc}
b_0 & b_0 & b_0 & \cdots & b_0 \\[1mm]
b_0 & b_1 & b_1 & \cdots & b_1 \\[1mm]
b_0 & b_1 & b_2 & \cdots & b_2 \\[1mm]
\vdots & & & & \vdots \\[1mm]
b_0 & b_1 & b_2 & \cdots & b_I
\ea\right]~.
\eq
We have
\beq \label{b2}
LM=\left[\ba{ccccc}
b_0 & b_0 & b_0 & & b_0 \\[1mm]
0 & b_1-b_0 & b_1-b_0 & & b_1-b_0 \\[1mm]
& & & & \\[1mm]
& & & & \\[1mm]
& & & b_{I-1}-b_{I-2} & b_{I-1}-b_{I-2} \\[1mm]
0 & & & 0 & b_I-b_{I-1}
\ea\right]
\eq
(upper triangular matrix with $(LM)_{ij}=b_i-b_{i-1}$, $j=i,i+1,...,I$ for $i=0,1,...,I$ and $b_{-1}$ defined 0). Next, when $U$ is the $(I+1)\times(I+1)$ upper triangular matrix
\beq \label{b3}
U=\left[\ba{ccccc}
1 & c_0 & 0 & & 0 \\[1mm]
0 & 1 & c_1 \\[1mm]
& & & & \\[1mm]
& & & & \\[1mm]
& & & 1 & c_{I-1} \\[1mm]
0 & & & 0 & 1
\ea\right]~,~~~~~~c_i={-}\,\frac{b_i-b_{i-1}}{b_{i+1}-b_i}~,
\eq
(bidiagonal matrix with $1$'s on the main diagonal and $c_i$, $i=0,1,...,I-1\,$, on the first upper codiagonal), we have
\beq \label{b4}
ULM=\left[\ba{cccc}
b_0-b_{-1} & 0 & & 0 \\[1mm]
0 & b_1-b_0 \\[1mm]
& & & \\[1mm]
& & & 0 \\[1mm]
0 & & 0 & b_I-b_{I-1}
\ea\right]=D
\eq
(diagonal matrix with diagonal elements $b_i-b_{i-1}$, $i=0,1,...,I$). Therefore, $D^{-1}ULM=I$, and Eq.~(\ref{e52n}) follows on computing $D^{-1}U$ from Eqs.~(\ref{b3}, \ref{b4}).

We next determine the inverse of the matirx
\beq \label{b5}
(B_{n+2,s}^m)_{s,n=|m|(2)N_m}
\eq
that occurs in Eq.~(\ref{e65}), also see Eq.~(\ref{e61}). With
\beq \label{b6}
s=|m|+2u~,~~~~~~n=|m|+2k~,~~~~~~u,k=0,1,...,K~,
\eq
where we have set $K=\tfrac12\,(N_m-|m|)$, the matrix in Eq.~(\ref{b5}) assumes the form
\beq \label{b7}
C=(C_{uk})_{u,k=0,1,...,K}~,
\eq
where
\beq \label{b8}
C_{uk}=\left\{\ba{llll}
4(|m|+2u+1)(|m|+k+u+2)(k+1-u) & \!\!, & ~~~k\geq u & \!\!, \\[2mm]
0 & \!\!, & ~~~k<u & \!\!,
\ea\right.
\eq
for integer $u$, $k$, with $0\leq u,k\leq K$. Thus $C=D_1E$, where $D_1$ is the $(K+1)\times(K+1)$ diagonal matrix with diagonal elements $8(|m|+2u+1)$, $u=0,1,...,K\,$, and $E$ is the upper triangular matrix with entries
\beq \label{b9}
E_{uk}=\tfrac12\,(|m|+k+u+2)(k+1-u)=\sum_{j=u}^k\,(\tfrac12\,|m|+j+1)
\eq
for $0\leq u\leq k\leq K$. Setting $d_j=\frac12\,|m|+j+1$ for $j=0,1,...,K\,$, and letting $U_1$, $U_2$ the two bidiagonal upper triangular matrices, given by
\beq \label{b10}
\left[\ba{ccccc}
1 & -1 & 0 & & 0 \\[1mm]
0 & 1 & -1 \\[1mm]
& & & & \\[1mm]
& & & 1 & -1 \\[1mm]
0 & & & 0 & 1
\ea\right]~,~~~~~~\left[\ba{ccccc}
1 & -d_0/d_1 & 0 & & 0 \\[1mm]
0 & 1 & -d_1/d_2 \\[1mm]
& & & & \\[1mm]
& & & 1 & -d_{K-1}/d_K \\[1mm]
0 & & & 0 & 1
\ea\right]
\eq
respectively, it is seen that
\beq \label{b11}
U_2(U_1E) = U_2\,\left[\ba{cccc}
d_0 & d_0 & & d_0 \\[1mm]
0 & d_1 & & d_1 \\[1mm]
& & & \\[1mm]
0 & & 0 & d_K
\ea\right] = \left[\ba{cccc}
d_0 & 0 & & 0 \\[1mm]
0 & d_1 \\[1mm]
& & & \\[1mm]
& & & 0 \\[1mm]
0 & & 0 & d_K
\ea\right]=D_2~.
\eq
Therefore, from $E=D_1^{-1}C$ and Eq.~(\ref{b11}) we get
\beq \label{b12}
C^{-1}=D_2^{-1}U_2U_1D_1^{-1}~.
\eq
Finally, we compute $U_2U_1$ from Eq.~(\ref{b10}) as the upper triangular matrix with $1$'s on the main diagonal, $-1-d_u/d_{u+1}$, $u=0,1,...,K-1\,$, on the first upper codiagonal, $d_u/d_{u+1}$, $u=0,1,...,K-2$ on the second upper codiagonal and 0 elsewhere. Hence, with the definitions of $D_1$ and $D_2$ as given, we have that $C^{-1}$ is an upper triangular tridiagonal matrix whose non-zero entries are given for $k,k'=0,1,...,K$ by
\begin{eqnarray} \label{b14}
(C^{-1})_{kk'} & = & \frac{(U_2\,U_1)_{kk'}}{4(|m|+2k+2)(|m|+2k'+1)} \nonumber \\[3.5mm]
& = & \left\{\ba{llll}
\dfrac{1}{4(|m|+2k+2)(|m|+2k+1)} & \!\!, & ~~~k=k' & \!\!, \\[4mm]
\dfrac{-1}{2(|m|+2k+2)(|m|+2k+4)} & \!\!, & ~~~k=k'-1 & \!\!, \\[4mm]
\dfrac{1}{4(|m|+2k+4)(|m|+2k+5)} & \!\!, & ~~~k=k'-2 & \!\!.
\ea\right. \nonumber \\
\mbox{}
\end{eqnarray}
From this the result of Eq.~(\ref{e67}) can be obtained by restoring the index $n'=|m|+2k$.\\[6mm]
{\bf \Large Acknowledgement}\\[2mm]
It is a pleasure to thank Professor J. Braat for his constant interest and feedback during the development of the results of this paper, in addition to providing technical assistance of various sorts. Furthermore, a comment of Professor R. Aarts, yielding an enhancement of the result of Section 3, is appreciated. \\[1mm]

\end{document}

%% file: layout.tex
\newcommand{\os}{\underline}
\newcommand{\mc}{\multicolumn}

\newcommand{\ba}{\begin{array}}
\newcommand{\ea}{\end{array}}

\newcommand{\vart}{\vartheta}
\newcommand{\varp}{\varphi}
\newcommand{\eps}{\varepsilon}

\newcommand{\il}{\int\limits}
\newenvironment{inspring}[1]%
{\begin{list}{}{\setlength{\rightmargin}{0cm}
                \setlength{\listparindent}{0cm}
                \settowidth{\labelwidth}{\mbox{#1}}
                \setlength{\leftmargin}{1.1\labelwidth}
                \setlength{\labelsep}{.1\labelwidth}}}%
{\end{list}}
\newcommand{\ITEM}[1]{\item[#1\hfill]}
\newcommand{\bi}[1]{\begin{inspring}{#1}}
\newcommand{\ei}{\end{inspring}}

\newcommand{\dsum}{\displaystyle \sum}

\newcommand{\beq}{\begin{equation}}
\newcommand{\eq}{\end{equation}}
\catcode`\@=11

\font\tenmsa=msam10 \font\sevenmsa=msam7 \font\fivemsa=msam5
\font\tenmsb=msbm10 \font\sevenmsb=msbm7 \font\fivemsb=msbm5
\newfam\msafam
\newfam\msbfam
\textfont\msafam=\tenmsa  \scriptfont\msafam=\sevenmsa
  \scriptscriptfont\msafam=\fivemsa
\textfont\msbfam=\tenmsb  \scriptfont\msbfam=\sevenmsb
  \scriptscriptfont\msbfam=\fivemsb

\def\Bbb{\ifmmode\let\next\Bbb@\else
 \def\next{\errmessage{Use \string\Bbb\space only in math mode}}\fi\next}
\def\Bbb@#1{{\Bbb@@{#1}}}
\def\Bbb@@#1{\fam\msbfam#1}